\documentclass[prl,twocolumn,superscriptaddress]{revtex4-2}
\usepackage{amsmath,amssymb,mathrsfs}
\usepackage{graphicx}
\usepackage{colortbl}
\usepackage{CJK}
\begin{document}
\begin{CJK*}{UTF8}{bsmi}
\title{Signatures of quantum criticality in the complex inverse temperature plane}

\author{Yang Liu (\CJKfamily{gbsn}刘洋)}
\affiliation{Key Laboratory of Polar Materials and Devices (MOE), School of Physics and Electronic Science, East China Normal University, Shanghai 200241, China}

\author{Songtai Lv (\CJKfamily{gbsn}吕松泰)}
\affiliation{Key Laboratory of Polar Materials and Devices (MOE), School of Physics and Electronic Science, East China Normal University, Shanghai 200241, China}

\author{Yang Yang (\CJKfamily{gbsn}杨洋)}
\affiliation{Key Laboratory of Polar Materials and Devices (MOE), School of Physics and Electronic Science, East China Normal University, Shanghai 200241, China}

\author{Haiyuan Zou  (\CJKfamily{gbsn}邹海源)}
\altaffiliation{hyzou@phy.ecnu.edu.cn}
\affiliation{Key Laboratory of Polar Materials and Devices (MOE), School of Physics and Electronic Science, East China Normal University, Shanghai 200241, China}

\begin{abstract}

Concepts of the complex partition functions and the Fisher zeros provide intrinsic statistical mechanisms for finite temperature and real time dynamical phase transitions. We extend the utility of these complexifications to quantum phase transitions. We exactly identify different Fisher zeros on lines or closed curves and elucidate their correspondence with domain-wall excitations or confined mesons for the one-dimensional transverse field Ising model. The crossover behavior of the Fisher zeros provides a fascinating picture for criticality near the quantum phase transition, where the excitation energy scales are quantitatively determined. We further confirm our results by tensor network calculations and demonstrate a clear signal of deconfined meson excitations from the disruption of the closed zero curves. Our results unambiguously show significant features of Fisher zeros for a quantum phase transition and open up a new route to explore quantum criticality.
\vspace{0.2cm}

\end{abstract}

\maketitle
\end{CJK*}
Quantum phase transition (QPT) takes place in between different ground states at zero temperature where thermal fluctuations are absent~\cite{Sachdevbook,Vojta_2003}. It is an essential topic in condensed matter physics and provides crucial ingredients for the understanding of many emerged properties in strongly correlated materials. Although QPT appears only at zero temperature, especially for a typical one-dimensional (1D) system where the extension of the transition to non-zero temperatures is forbidden, the influence of QPT provides various critical behaviors in an expansive finite temperature vicinity of the associated quantum critical point (QCP). Therefore, besides the typical scenario of projecting to the ground states at extremely large inverse temperature ($\beta$) and probing the excitations from different correlations, finding direct statistical approaches may provide new route to explore the possible remaining quintessence of quantum criticality at finite temperatures due to the interplay between the quantum and thermal fluctuations.

An important statistical mechanism is the analytic continuation to complex partition functions, pioneered by Lee and Yang~\cite{LeeYang1,LeeYang2}, who first gave the circle theorem of the grand canonical partition function zeros for the complex fugacity. This concept has been realized experimentally~\cite{zero_exp1,zero_exp2}. Fisher's extension of zeros for canonical partition function in the complex $\beta$ plane~\cite{Fisher1965} has triggered tremendous studies on this complexification concept in both equilibrium and nonequilibrium systems. For an equilibrium system with a finite temperature phase transition, Fisher zeros approach the real $\beta$ axis in the thermodynamic limit~\cite{Saarloos1984,zero_review} and the approaching behaviours can characterize different kind of phase transitions, such as the Kosterlitz-Thouless transition~\cite{Zou2014PRD}. Base on the fact that the zeros closest to the real $\beta$ axis can be seen as a ``gate" controlling the complex renormalization group (RG) flows~\cite{RGflow2010,Zou2011PRD,Liu2011PRD}, in the special case of a finite temperature phase transition, zeros on the real $\beta$ terminate the RG flows to different phases and can form an effective phase diagram in the complex $\beta$ plane~\cite{Sankhya2022PRR}. In a nonequilibrium quenched system, the Fisher zeros of the boundary partition function (or the Loschmidt echo) can be defined and approaching of these zeros to the real time axis signifies a dynamical phase transition~\cite{DynamicalPT2013PRL,DynamicalPT2014PRB,reviewDPT2016,Heyl_2018review}. These efforts of complexification on finite temperature or real time (infinite temperature) phase transitions motivate one to extend the mechanism for QPTs.

In a 1D quantum system without finite temperature transitions, the Fisher zeros are distributed on the complex $\beta$ without touching the real $\beta$ axis. Investigating QPTs from the complex $\beta$ ($\beta=\beta_r+i\beta_i$) seems a marginal problem. However, $\beta_i$ suggests an oscillating behavior with the energy scale of the excitations in a partition function. Therefore, information of the complex $\beta$ provides ingredients to study the interplay between thermal and quantum fluctuations. Specifically, Locations and shapes of Fisher zeros can intimate low temperature excitation features of QPT. For example, similar with a close zero curve cutting the real $\beta$ axis can separate different phases at a finite temperature transition, we argue that this picture can be naturally generalized to QPTs where the chief distinctive feature is that the close zero curves in the complex $\beta$ plane isolate excitations. 

\begin{figure}[h]
\includegraphics[width=0.48\textwidth]{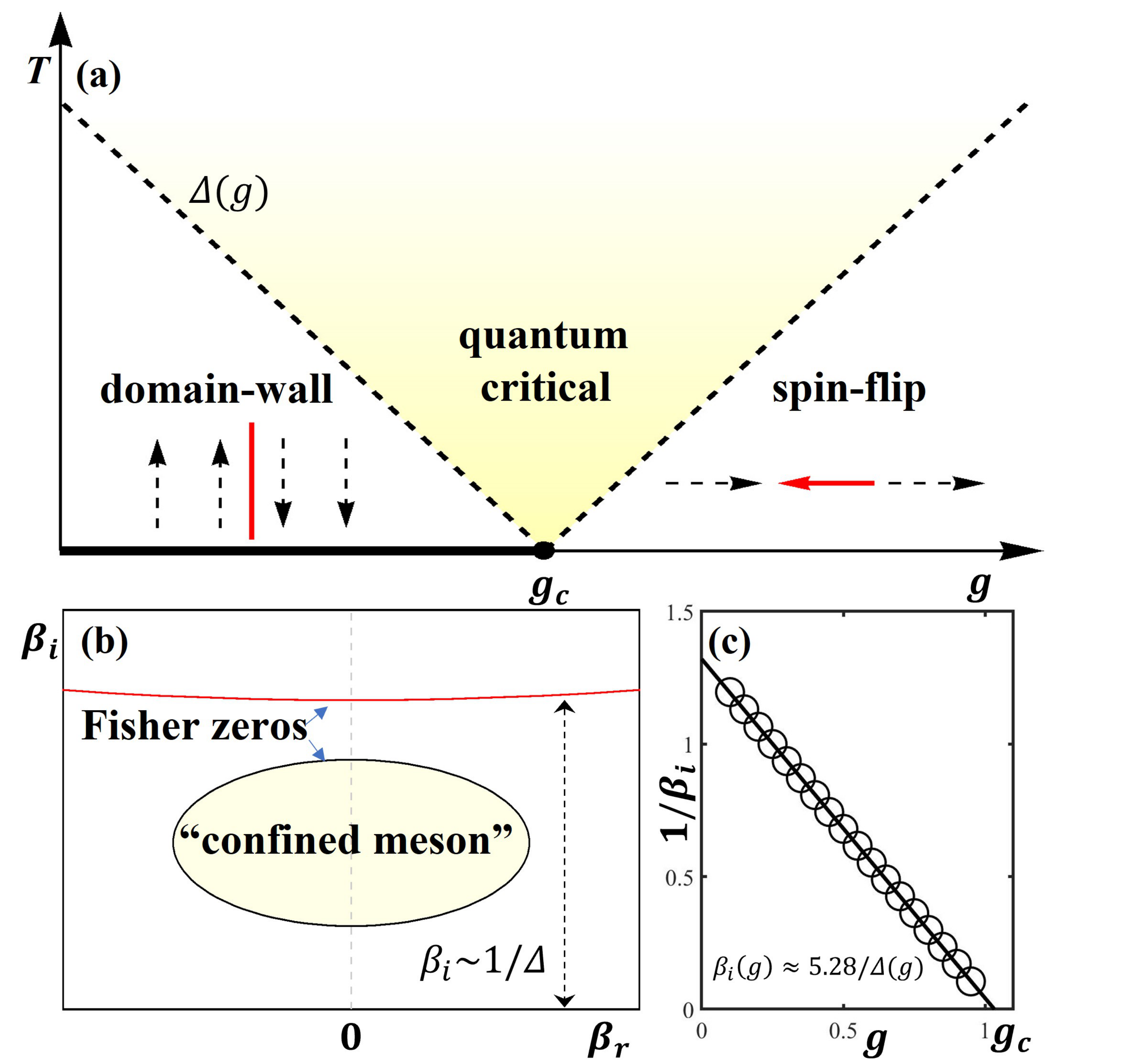}
\caption{Quantum criticality and Fisher zeros. (a) Schematic phase diagram of the 1DTFIM. (b) The Fisher zeros in the complex $\beta$ plane: the imaginary part of the zero line provides an energy scale for the gap of the domain-wall excitations. Meanwhile, the closed zero curve confines the meson excitations from the 1DTFIM ground state. Quantum criticality is signified by the competition between these two structures. (c) At a relatively large $\beta_r=8$, the imaginary part of the zero inversely proportional to $g$ with a fitting form $1/\beta_i=-1.28(g-1.03)$, confirms the relation between $\beta_i$ and the domain-wall excitation gap as $\beta_i\approx5.28/\Delta(g)$. [Note that the solution of Fisher zeros is obtained from Eq.~(4)].}
\label{fig:F1qcp}
\end{figure}
Here, we confirm these above pictures of the Fisher zeros and establish the connections between these zeros and QPT in the 1D transverse field Ising model (1DTFIM) [Fig.~\ref{fig:F1qcp}(a)]. The domain-wall excitations and the confined mesons can be represented by the Fisher zero lines and closed curves in the complex plane, respectively [Fig.~\ref{fig:F1qcp}(b)]. Specifically, the gap of the domain-wall excitations can be constructed from information of the zero lines precisely [Fig.~\ref{fig:F1qcp}(c)]. Changing coupling leads to crossover of the lines and the close curves till at the QCP where the lines disappear and the whole large $\beta$ region are connected. We verify the excitation energy gaps through the oscillation of a Loschmidt-Echo-like quantity-the norm of the partition function. Our picture of the Fisher zeros is confirmed by tensor network calculation at finite volumes. Remarkably, the confined meson picture is clearly illustrated by breaking of closed zero curves as a longitudinal field is applied at the QCP, corresponding to the generation of exotic $E_8$ quasi-particles. Our results on using the Fisher zeros structure and the complex partition functions to signify the quantum criticality provide a new framework to explore QPTs.

\begin{figure}[t]
\includegraphics[width=0.5\textwidth]{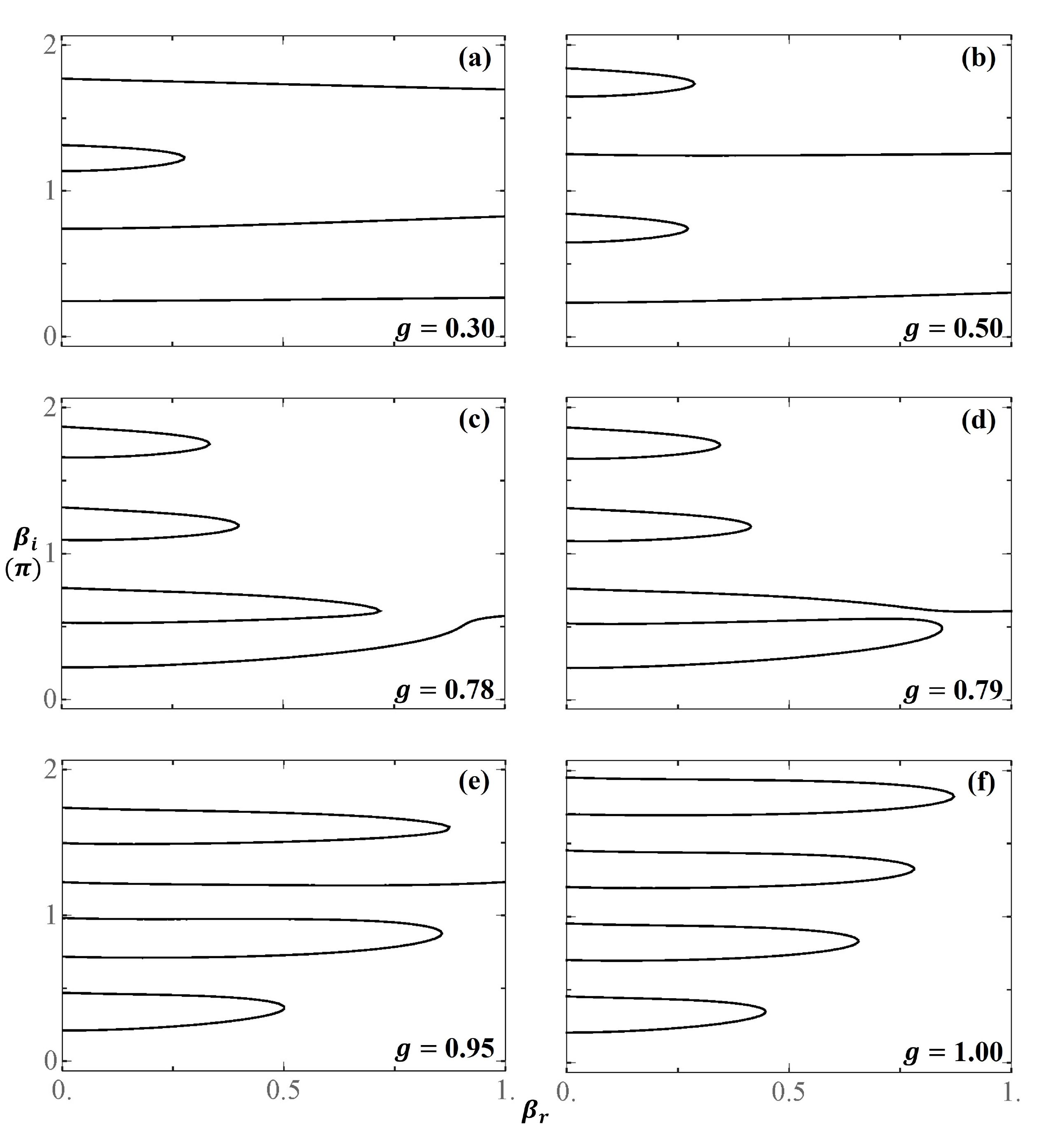}
\caption{Distribution patterns of the Fisher zero lines and the closed curves at different $g$. The zero lines move upward as $g$ increases. (a) At $g=0.3$, the zero lines are the dominant structure. (b) At $g=0.5$, the zero lines and the closed curves are alternatively appears. (c-d) The crossover between the lowest zeros line and the close curve above the real $\beta$ axis happens in between $g=0.78$ and $g=0.79$. (e) At $g=0.95$, the close curves are the dominant structure. (f) After more crossovers, the zero lines disappear at the QCP.}
\label{fig:F2zeros}
\end{figure}
We consider the exact solvable 1DTFIM--a paradigmatic model for QPT--whose Hamiltonian is given by
\begin{equation}
H=-\sum_{j=1}^L\sigma_j^z\cdot\sigma_{j+1}^z-g\sigma_j^x,
\label{eq:TFIC}
\end{equation}
where $\sigma_j^x$ or $\sigma_j^z$ are the Pauli matrix operators on site $j$. At the critical magnetic field coupling $g=g_c=1$, a QPT between the ferromagnetic order and the paramagnetic phase appears only at the zero temperature in the thermodynamic limit. 

Explicit statistical properties can be obtained from the quantum model partition function,
\begin{equation}
Z=\mathrm{Tr}\hspace{1pt} e^{-\beta H},
\label{eq:partf}
\end{equation}
where $\beta=1/k_BT$ is the inverse temperature and the trace is over all states of the system. The Fisher zeros are particular $\beta$s in the complex plane where the partition function is vanished. 
The exact solution of the partition function for the 1DTFIM is expressed as 
\begin{eqnarray}
\nonumber
Z&=&\frac{1}{2}[\prod_{k=1}^L(2\cosh\beta\epsilon_{2k})+\prod_{k=1}^L(2\sinh\beta\epsilon_{2k})\\
&+&\prod_{k=1}^L(2\cosh\beta\epsilon_{2k-1})+\prod_{k=1}^L(2\sinh\beta\epsilon_{2k-1})]
\label{eq:exact}
\end{eqnarray}
where $2\epsilon_k$ is the single particle spectrum with $\epsilon_k=[1+g^2-2g\cos(\pi k/L)]^{1/2}$. This is derived from the Onsager/Kaufman solutions~\cite{Onsager1944,Kaufman1949} of the two-dimensional classical Ising model by using the mapping between the $d$-dimensional quantum spin model and the $d$$+$$1$-dimensional classical statistical model~\cite{Suzuki1976}. Eq.~(\ref{eq:exact}) can be expressed as terms with the product of $\prod_{k=1}^L(2\cosh\beta\epsilon_{2k/2k-1})$ [term (I)] and $\prod_{k=1}^L(\tanh\beta\epsilon_{2k/2k-1})+1$ [term (II)]. Conventionally, thermodynamic limit properties of the 1DTFIM at zero temperature can be well understood by ignoring the term (II)~\cite{PFEUTY197079}. However, this approximation has limitations for digging the detail information of the excitations at finite temperature~\cite{Biaoczyk2021}.

Although the quantum-classical mapping exists, the thermal properties implied by their complex partition functions are not the same. For the 2D classical Ising model with a finite temperature phase transition, the Fisher zeros approach the real $\beta$ axis in the thermodynamic limit. In contrast, zeros stay away from the real axis for the 1D quantum case. For example, at $g=0$, zeros are formed in lines parallel with the real $\beta$ axis at $g=0$~\cite{Jones1966}. However, a detail structure of the Fisher zeros in general cases of the 1DTFIM and their relations to quantum criticality of the model are seldom discussed topics.

The Fisher zeros contributed from term (I) form lines only on the imaginary $\beta$ axis and varying the coupling $g$ does not affect the structure of these zeros. Therefore, we reconsidered the term (II) for questing the thermal properties. The condition for $Z=0$ then can be expressed as  
$\sum_{k=1}^L\log|\tanh\beta\epsilon_{2k/2k-1}|^2=0$ after few steps of reformation. In the thermodynamic limit ($L\rightarrow\infty$), it can be expressed as a continuous expression,
\begin{equation}
\int_0^\pi\log|\tanh\beta\epsilon_q|^2dq=0,
\label{eq:intg}
\end{equation} 
where $\epsilon_q=(1+g^2-2g\cos q)^{1/2}$. Note that the Fisher zeros solution of Eq.~(\ref{eq:intg}) have both the symmetry $\beta_r\rightarrow -\beta_r$ and $\beta_i\rightarrow -\beta_i$, we therefore consider the cases with $\beta_r>0$ and $\beta_i>0$, shown in Fig.~\ref{fig:F2zeros}.

A clear crossover behavior between the Fisher zero lines and the closed zero curves arises as $g$ increases, with the trend of upward moving of lines and increased numbers of curves. At small $g$ [e.g. Fig~\ref{fig:F2zeros}(a)], the Fisher zero lines separate the complex plane into different sectors and inhibit the spreading of complex RG flow to the whole large complex $\beta$ region. Meanwhile, small numbers of the closed zero curves appear at large $\beta_i$. At $g=0.5$ [Fig~\ref{fig:F2zeros}(b)], the zero lines and the closed curves distribute alternatively in the direction of $\beta_i$. At $g\sim 0.8$ [Fig~\ref{fig:F2zeros}(c,d)], the crossover happens between the lowest line and the first close curve from the real $\beta$ axis. Afterwards, the lowest line rises rapidly, crosses more curves [Fig~\ref{fig:F2zeros}(e)], until fades away to the infinite $\beta$ region at the QCP $g=1$ [Fig~\ref{fig:F2zeros}(f)]. Drastic change of the lowest zero line near $g=1$ implies crossover behavior at finite temperature of thermal quantities and quantum criticality near the QCP, where the Fisher zero lines disappear and the complex RG flow can pass to the whole large complex $\beta$ region. Note that for $g>1$, due to the duality of the model, the zeros rescale in a plane with the unit $\beta/g$ are conformally the same with those in the $\beta$ plane for $1/g$.

Our results of the complex partition functions not only give a vivid crossover picture of the Fisher zeros' structure, but also suggest quantitative analyses for the quantum criticality of the excitations. Considering the partition function with the form $\sum a_i\exp(-\beta E_i)$, [e.g., $E_0$($E_1$) is the ground state (first excited state) energy], the norm $|Z(\beta)|^2$ contains the term $\exp(\pm i\beta_i\Delta)$, where $\Delta=E_1-E_0$. Therefore it provides a direct way to probe the excitation gaps. It is worthwhile emphasizing that this quantity is an analogy of Loschmidt Echo~\cite{Sun2006PRL} for the boundary partition functions in a dynamical phase transition and play a similar role as the second R\'enyi entropy~\cite{Karl2022}.

\begin{figure}[t]
\includegraphics[width=0.48\textwidth]{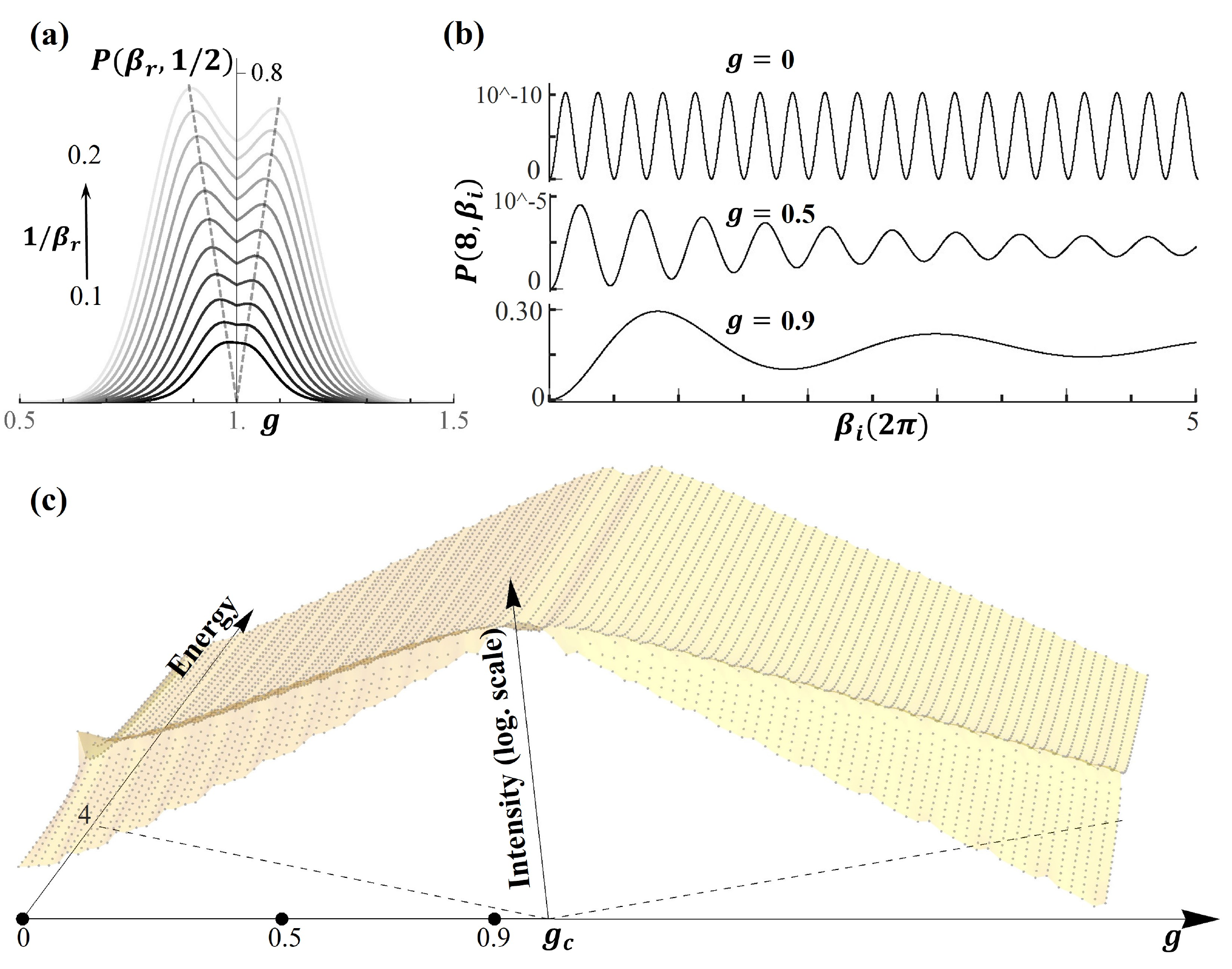}
\caption{Quantum critical behaviors from the relative norm of the complex partition function $P(\beta_r,\beta_i)$ of a finite 1DTFIM system with $L=64$ are shown. (a) at fixed $\beta_i=1/2$, $P(\beta_r, 1/2)$ as a function of $g$ shows a crossover behavior as the temperature $1/\beta_r$ increases from 0.1 to 0.2. The dashed lines show a linear scaling with the peaks. (b) At fixed large $\beta_r$, oscillation of $P$ as a function of $\beta_i$ at different $g$s indicates low energy excitations. The amplitude increases drastically near the QCP. The Fourier transform shows intensity peaks of excitations in panel (c), gives a clear signal of the low energy excitations.}
\label{fig:F3critical}
\end{figure}

We define the relative norm of partition function 
\begin{equation}
P(\beta_r,\beta_i)=1-\frac{|Z(\beta)|^2}{|Z(\beta_r)|^2},
\label{eq:relnorm}
\end{equation}
where $\beta=\beta_r+i\beta_i$ and $0\le P\le 1$ ($P=1$ at Fisher zeros and vanishes on the real $\beta$ axis). Results of $P$ for a finite system show crossovers with varied $\beta_r$ at a fixed energy scale ($1/\beta_i$) [Fig.~\ref{fig:F3critical}(a)] and oscillations with varied $\beta_i$ at a fixed temperature ($1/\beta_r$) [Fig.~\ref{fig:F3critical}(b)]. The former case is consistent with the crossover behavior of other thermal quantities, e.g., the specific heat~\cite{Wu2018PRB}. For the latter case, the increased amplitudes near QCP imply the quantification to characterize quantum fluctuation at large finite temperature, and the oscillating frequencies signify the excitation energies. The peaks in the corresponding spectrum of $P$ match the IDTFIM excitation energy gaps precisely [Fig.~\ref{fig:F3critical}(c)]. Note that other quantities, such as $\log(|Z(\beta)|)$ related with the free energy, can also be used to extract the energy spectrum.

The closed Fisher zero curves also provide new features of quantum criticality. Unlike the closed zero curves which touch the real $\beta$ axis to separate different real phases in a finite temperature phase transition, the closed curves in the complex $\beta$ plane are symmetry protected and represent confined structures in QPT. These structures are the effects due to the quantum fluctuations,  different from the kink confinement to form normal mesons~\cite{Coldea2010}. Correspondingly, the closed zero curves imply emergent meson excitations after symmetry breaking. For the 1DTFIM, closeness of the zero curves is robust against the changing of $g$. Adding additional $Z_2$ symmetry breaking longitudinal field can break the close curves and give rise to exotic meson excitations. This picture is highly consistent with the fact of the emerging $E_8$ meson excitations~\cite{ZAMOLODCHIKOV1989,Coldea2010,Wang2020PRB,Zou2021PRL} for the system with Hamiltonian 
\begin{equation}
H=-\sum_{j=1}^L(\sigma_j^z\cdot\sigma_{j+1}^z+g\sigma_j^x+h\sigma_j^z),
\label{eq:E8}
\end{equation}
where $g=g_c=1$. 
However, unlike the 1DTFIM, the partition function with a longitudinal field is still unclear. To confirm the confined/deconfined meson picture, numerical investigations of the Fisher zeros are demanded. 

The quantum-classical mapping implies that the partition function of 1D quantum spin model can be represented by a 2D lattice network structure. Specifically, in a tensor network language~\cite{ORUS2014117,TNreview1,TNreview2}, the imaginary time evolution $\exp(-\beta H)$ can be constructed as matrix product operators (MPOs) using the Trotter formula with $\beta=\tau N$ at large $N$. The partition function of the quantum system Eq.~(\ref{eq:TFIC}) or (\ref{eq:E8}) with length $L$ is equivalent to a $L\times N$ 2D classical system. We employ the HOTRG method~\cite{XieHOTRG} which had been efficiently applied to systems with the complex sign problems~\cite{Zou2014PRD} and approximate the MPOs into transfer matrices with a tunable tensor bond dimension $D_b$. In our calculation, $D_b$ is increased up to 40 for systems with the volume $64\times1024$ and convergent results of Fisher zeros are obtained. 

\begin{figure}[h]
\includegraphics[width=0.46\textwidth]{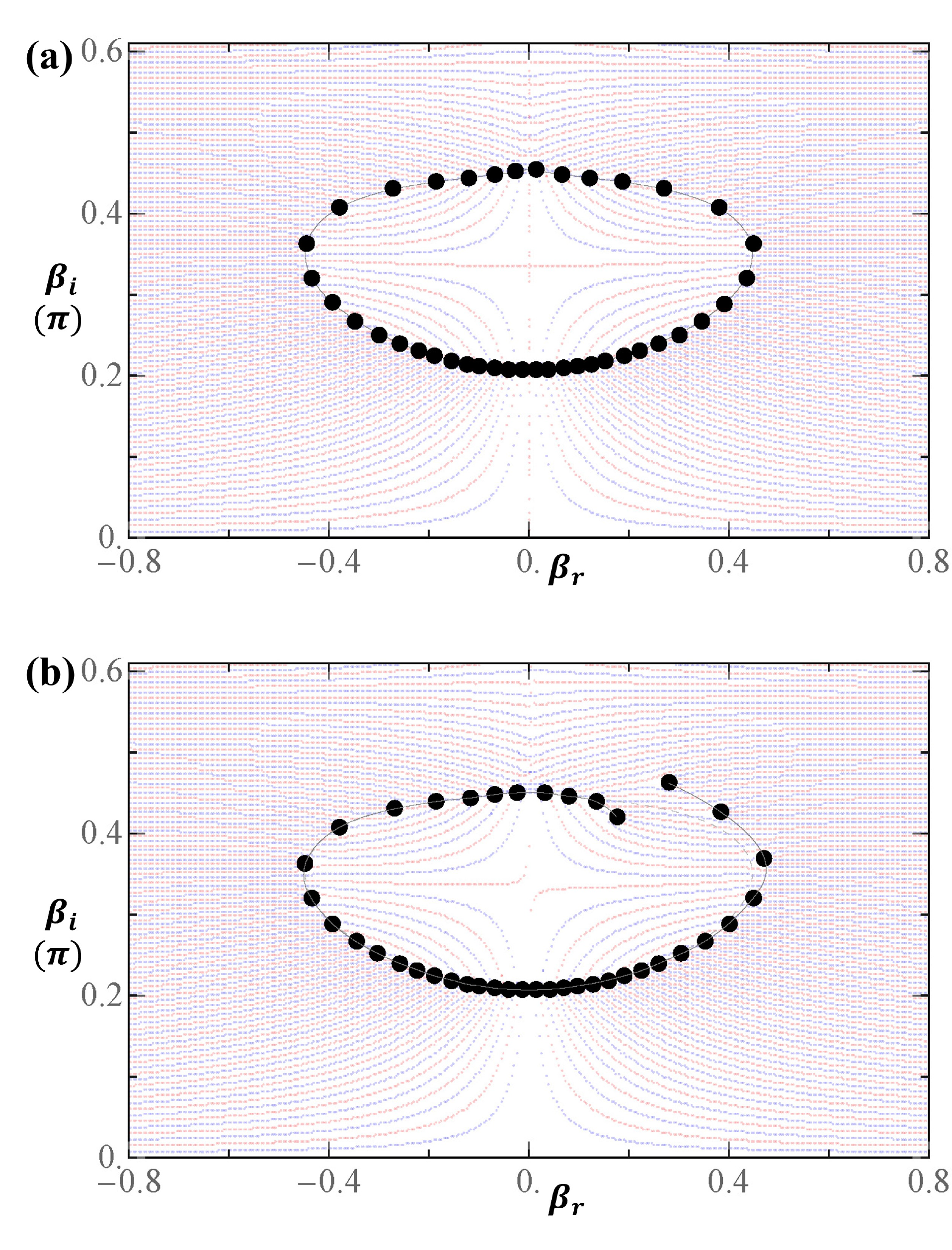}
\caption{Fisher zeros (black dots) in the ($\beta_r$,$\beta_i$) plane for the 1DTFIM at the QCP and the situation with a longitudinal field $h$ at finite $L=64$. Zeros are on the intersects of the solutions for the real (blue) and the imaginary (red) parts. (a) At $h=0$, the results is consistent with the thermodynamic limit closed zero curve (gray line), indicate ``confined" meson excitations. (b) At $h=0.02$, the closed curve shape is disrupted, which indicates the real meson excitations appeared in reality.}
\label{fig:F4e8zero}
\end{figure}

\begin{figure}[t]
\includegraphics[width=0.46\textwidth]{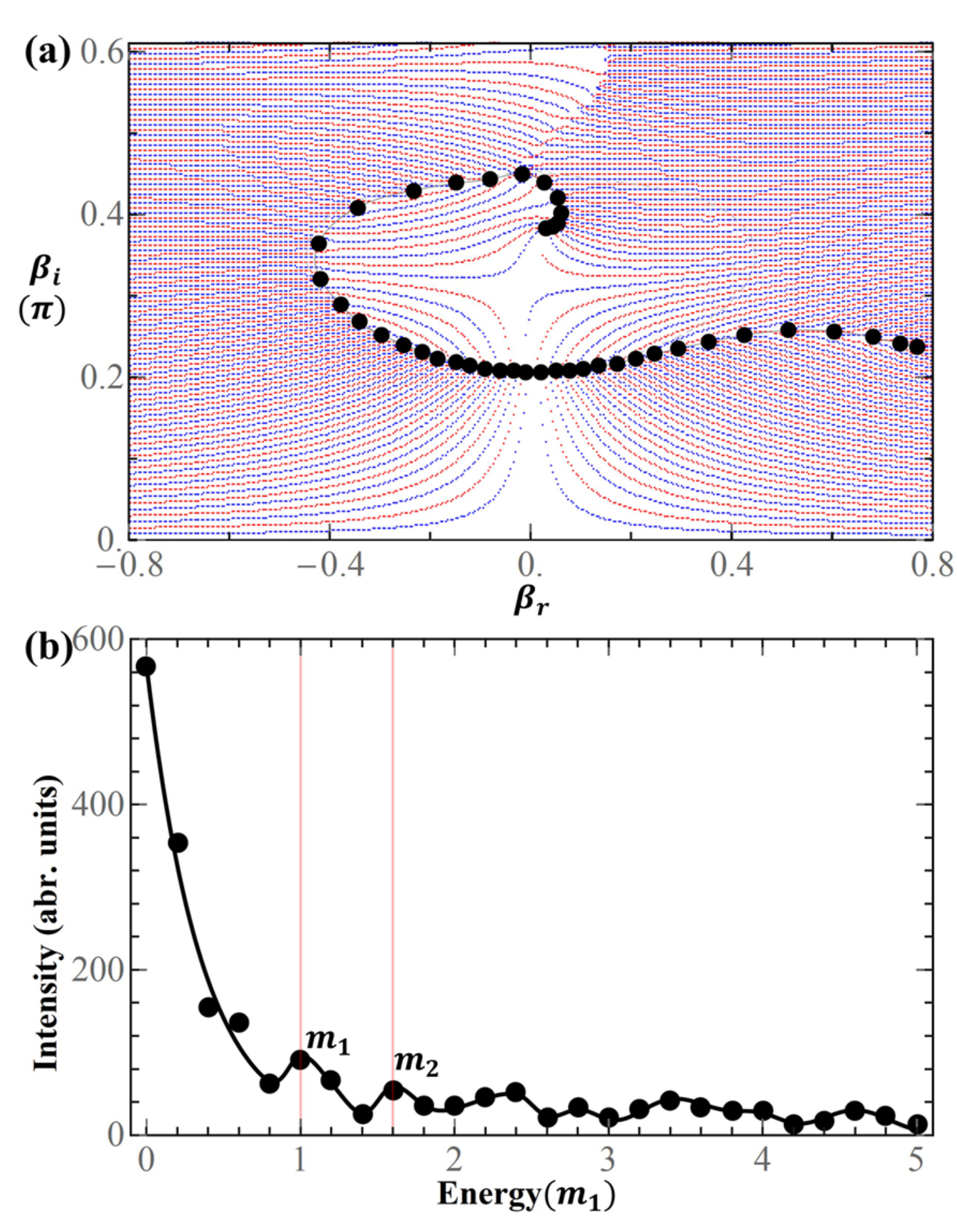}
\caption{Fisher zeros and energy spectrum at larger $h$. (a) The lowest Fisher zeros (black dots) in the ($\beta_r$,$\beta_i$) plane for the 1DTFIM with a longitudinal field $h=0.1$ at finite $L=64$. (b) The energy spectrum calculated from the complex partition function with $D_b=60$. The mass ratio between the first two mesons $m_2/m_1\approx 1.6$, close to the golden ratio 1.618.}
\label{fig:F5e8}
\end{figure}

For the 1DTFIM at the QCP without the longitudinal field $h$, the lowest Fisher zeros above the real $\beta$ axis for a finite system distribute densely on the thermodynamic limit closed zero curve [Fig.~\ref{fig:F4e8zero}(a)]. These zeros are disrupted by a large gap when a small longitudinal field $h$ is applied [Fig.~\ref{fig:F4e8zero}(b)]. It implies breaking of the closed zero curve in the thermodynamic limit and meson excitations flowing out to the large $\beta$ region. As $h$ increases, the zero curve is further disrupted [Fig.~\ref{fig:F5e8}(a)]. We also calculate the energy spectrum from the complex partition function by using $\log(|Z(\beta)|)$. The mass ratio between the first two mesons is closed to the golden ratio in the $E_8$ mesons [Fig.~\ref{fig:F5e8}(b)]. These phenomena in complex $\beta$ plane unambiguously confirms the picture that the close zero curves confine the meson excitations for the 1DTFIM.  

Our results can provide useful experimental and theoretical insights. Firstly, the stability of the lowest closed zero curve above the real $\beta$ axis at $g>0.8$~[Fig.~\ref{fig:F2zeros}(d)] implies that the curve breaking picture under the additional field is a generic feature and suggests a robust region of meson excitations in the vicinity of QCP. This is consistent with the experimental observation on quasi-particles in a large range of coupling $g$ tuned by the applied magnetic field~\cite{Wang2020t}. The crossover between the lowest zero line and closed zero curve at $g\sim 0.8$ gives an energy scale of the competition between domain-wall excitations and confined mesons and suggests future inelastic neutron scattering or terahertz measurements on the decay of mesons (or meson melting~\cite{Karl2022}) at $g$(or $1/g$)$\sim 0.8$. Secondly, the relative norm of partition function $P$ is an alternative quantity to classify the excitations in comparison with the space-time correlation calculation from the ground states with a volume-law-increased quantum entanglement. Furthermore, by the aid of our current numerical approach and other tensor network contraction methods developed for higher dimensional statistical models~\cite{XieHOTRG,XieCPL2022,Yangarxiv2022}, our framework can be extended to Lee-Yang zero studies~\cite{Liu2006LeeYang} and other phase transition cases, such as symmetry protected phase transitions~\cite{Wen2013,Zou2019PRL,Zou2020PRB}, conditional 1D finite temperature phase transition~\cite{FiniteT2016,ZhangPRL2021}, superradiant phase transition~\cite{Luo2023} and deconfined QPT~\cite{Senthil2004}. Lastly, although it is a difficult task in general, criticality from Fisher zeros is experimental detectable through time-dependent quantum coherence measurement by mapping the partition function to $\mathrm{Tr}[\exp(-\beta H-itH_I)]$ with the probe-bath interaction $H_I=H$~\cite{zero_exp1,Wei2014}.

In summary, we demonstrate the exact Fisher zeros and their connections to quantum criticality for the prototypical 1DTFIM in the thermodynamic limit. The crossover between two different kinds of Fisher zeros provides an excitation competing picture near the QCP. The energy scale of domain-well suggested by the Fisher zero lines is quantified by the oscillating properties of the partition function norm, while the confined meson characterized by the closed Fisher zero curves is justified by the curve disruption behavior under a symmetry breaking longitudinal perturbation. Our statistical complex finite temperature analyses suggest an unexplored perspective for various kind of QPTs in general.

We thank Fernando G\'omez-Ruiz and Johannes Knaute for helpful discussions. This work is supported by National Natural Science Foundation of China Grant No.~12274126. 


\begin{thebibliography}{47}

\bibitem{Sachdevbook}
S. Sachdev,
{\it Quantum Phase Transitions} (Cambridge University Press, Cambridge, England, 2011).


\bibitem{Vojta_2003}
M. Vojta,
\newblock Reports on Progress in Physics \textbf{66}, 2069 (2003).

\bibitem{LeeYang1}
C. N. Yang and T. D. Lee,
\newblock Phys. Rev. \textbf{87}, 404 (1952).

\bibitem{LeeYang2}
T. D. Lee and C. N. Yang,
\newblock Phys. Rev. \textbf{87}, 410 (1952).

\bibitem{zero_exp1}
X. Peng, H. Zhou, B.-B. Wei, J. Cui, J. Du, and R.-B.Liu,
\newblock Phys. Rev. Lett. \textbf{114}, 010601 (2015).

\bibitem{zero_exp2}
K. Brandner, V. F. Maisi, J. P. Pekola, J. P. Garrahan, and C. Flindt,
\newblock Phys. Rev. Lett. \textbf{118}, 180601 (2017).

\bibitem{Fisher1965}
M. E. Fisher, 
{\it Statistical Physics, Weak Interactions, Field Theory, Lectures in Theoretical Physics Vol. VIIC} (University of Colorado Press, Boulder, 1965).

\bibitem{Saarloos1984}
W. van Saarloos and D. A. Kurtze,
\newblock Journal of Physics A: Mathematical and General \textbf{17}, 1301 (1984).

\bibitem{zero_review}
I. Bena, M. Droz, and A. Lipowski,
\newblock International Journal of Modern Physics B \textbf{19}, 4269 (2005).

\bibitem{Zou2014PRD}
A. Denbleyker, Y. Liu, Y. Meurice, M. P. Qin, T. Xiang, Z. Y. Xie, J. F. Yu, and H. Zou,
\newblock Phys. Rev. D \textbf{89}, 016008 (2014).

\bibitem{RGflow2010}
A. Denbleyker, D. Du, Y. Liu, Y. Meurice, and H. Zou,
\newblock Phys. Rev. Lett. \textbf{104}, 251601 (2010).

\bibitem{Zou2011PRD}
Y. Meurice and H. Zou,
\newblock Phys. Rev. D \textbf{83}, 056009 (2011).

\bibitem{Liu2011PRD}
Y. Liu and Y. Meurice,
\newblock Phys. Rev. D \textbf{83}, 096008 (2011).

\bibitem{Sankhya2022PRR}
S. Basu, D. P. Arovas, S. Gopalakrishnan, C. A. Hooley, and V. Oganesyan,
\newblock Phys. Rev. Research \textbf{4}, 013018 (2022).

\bibitem{DynamicalPT2013PRL}
M. Heyl, A. Polkovnikov, and S. Kehrein,
\newblock Phys. Rev. Lett. \textbf{110}, 135704 (2013).

\bibitem{DynamicalPT2014PRB}
F. Andraschko and J. Sirker,
\newblock Phys. Rev. B \textbf{89}, 125120 (2014).

\bibitem{reviewDPT2016}
A. Zvyagin,
\newblock Low Temperature Physics \textbf{42}, 971 (2016).

\bibitem{Heyl_2018review}
M. Heyl,
\newblock Reports on Progress in Physics \textbf{81}, 054001 (2018).

\bibitem{Onsager1944}
L. Onsager,
\newblock Phys. Rev. \textbf{65}, 117 (1944).

\bibitem{Kaufman1949}
B. Kaufman,
\newblock Phys. Rev. \textbf{76}, 1232 (1949).

\bibitem{Suzuki1976}
M. Suzuki,
\newblock Progress of Theoretical Physics \textbf{56}, 1454 (1976).

\bibitem{PFEUTY197079}
P. Pfeuty,
\newblock Annals of Physics \textbf{57}, 79 (1970).

\bibitem{Biaoczyk2021}
M. Bia{\l}o{\'{n}}czyk, F. G{\'{o}}mez-Ruiz and A del Campo,
\newblock SciPost Phys. \textbf{11}, 013 (2021).

\bibitem{Jones1966}
G. Jones,
\newblock Journal of Mathematical Physics \textbf{7}, 2000 (1966).

\bibitem{Sun2006PRL}
H. T. Quan, Z. Song, X. F. Liu, P. Zanardi, and C. P. Sun,
\newblock Phys. Rev. Lett. \textbf{96}, 140604 (2006).

\bibitem{Karl2022}
M. C. Banuls, M. P. Heller, K. Jansen, J. Knaute, and V. Svensson,
\newblock arXiv:2206.10528 (2022).

\bibitem{Wu2018PRB}
J. Wu, L. Zhu, and Q. Si,
\newblock Phys. Rev. B \textbf{97}, 245127 (2018).

\bibitem{Coldea2010}
R. Coldea, D. Tennant, E. Wheeler, E. Wawrzynska, D. Prabhakaran, M. Telling, K. Habicht, P. Smeibidl, and K. Kiefer,
\newblock Science \textbf{327}, 177 (2010).

\bibitem{ZAMOLODCHIKOV1989}
A. Zamolodchikov,
\newblock International Journal of Modern Physics A \textbf{04}, 4235 (1989).

\bibitem{Wang2020PRB}
Z. Zhang, K. Amelin, X. Wang, H. Zou, J. Yang, U. Nagel, T. R\~o\~om, T. Dey, A. A. Nugroho, T. Lorenz, J. Wu, and Z. Wang,
\newblock Phys. Rev. B \textbf{101}, 220411 (2020).

\bibitem{Zou2021PRL}
H. Zou, Y. Cui, X. Wang, Z. Zhang, J. Yang, G. Xu, A. Okutani, M. Hagiwara, M. Matsuda, G. Wang, G. Mussardo, K. H\'ods\'agi, M. Kormos, Z. He, S. Kimura, R. Yu, W. Yu, J. Ma, and J.Wu,
\newblock Phys. Rev. Lett. \textbf{127}, 077201 (2021).

\bibitem{ORUS2014117}
R. Or\'us,
\newblock Annals of Physics \textbf{349}, 117 (2014).

\bibitem{TNreview1}
J. I. Cirac, D. P\'erez-Garc\'{\i}a, N. Schuch, and F. Verstraete,
\newblock Rev. Mod. Phys. \textbf{93}, 045003 (2021).

\bibitem{TNreview2}
Y. Meurice, R. Sakai, and J. Unmuth-Yockey,
\newblock Rev. Mod. Phys. \textbf{94}, 025005 (2022).

\bibitem{XieHOTRG}
Z. Y. Xie, J. Chen, M. P. Qin, J. W. Zhu, L. P. Yang, and T. Xiang,
\newblock Phys. Rev. B \textbf{86}, 045139 (2012).

\bibitem{Wang2020t}
K. Amelin, J. Engelmayer, J. Viirok, U. Nagel, T. R\~o\~om, T. Lorenz, and Z. Wang,
\newblock Phys. Rev. B \textbf{102}, 104431 (2020).

\bibitem{XieCPL2022}
X. F. Liu, Y. F. Fu, W. Q. Yu, J. F. Yu, and Z. Y. Xie,
\newblock Chinese Physics Letters \textbf{39}, 067502 (2022).

\bibitem{Yangarxiv2022}
L.-P. Yang, Y. F. Fu, Z. Y. Xie, and T. Xiang,
\newblock arXiv:2210.09896 (2022).

\bibitem{Liu2006LeeYang}
P. Tong and X. Liu,
\newblock Phys. Rev. Lett. \textbf{97}, 017201 (2006).

\bibitem{Wen2013}
X. Chen, Z.-C. Gu, Z.-X. Liu, and X.-G. Wen,
\newblock Phys. Rev. B \textbf{87}, 155114 (2013).

\bibitem{Zou2019PRL}
H. Zou, E. Zhao, X.-W. Guan, and W. V. Liu,
\newblock Phys. Rev. Lett. \textbf{122}, 180401 (2019).

\bibitem{Zou2020PRB}
Q. Zheng, X. Li, and H. Zou,
\newblock Phys. Rev. B \textbf{101}, 165131 (2020).

\bibitem{FiniteT2016}
V. Michal, I. Aleiner, B. Altshuler, and G. Shlyapnikov,
\newblock Proceedings of the National Academy of Sciences \textbf{113}, E4455 (2016).

\bibitem{ZhangPRL2021}
K. L. Zhang and Z. Song,
\newblock Phys. Rev. Lett. \textbf{126}, 116401 (2021).

\bibitem{Luo2023}
Y.-T. Yang and H.-G. Luo,
\newblock Chinese Physics Letters \textbf{40}, 020502 (2023).

\bibitem{Senthil2004}
T. Senthil, A. Vishwanath, L. Balents, S. Sachdev, and M. Fisher
\newblock Science \textbf{303}, 1490 (2004).

\bibitem{Wei2014}
B.-B. Wei, S.-W. Chen, H.-C. Po, and R.-B. Liu,
\newblock Scientific Reports \textbf{4}, 5202 (2014).

\end{thebibliography}

\end{document}